\documentclass[
reprint,
aps,
pra,
superscriptaddress,
longbibliography
]{revtex4-1}

\usepackage{amsmath}
\usepackage{graphicx}
\usepackage{hyperref}
\hypersetup{
  colorlinks   = true, 
  urlcolor     = blue, 
  linkcolor    = blue, 
  citecolor    = blue 
}

\begin{document}
\title{Spectral alignment of single-photon emitters in diamond using strain gradient}

\author{Smarak Maity}
\thanks{These authors contributed equally.}
\author{Linbo Shao}
\thanks{These authors contributed equally. \\shaolb@seas.harvard.edu }

\author{Young-Ik Sohn}

\author{Srujan Meesala}

\author{Bartholomeus Machielse}
\affiliation{John A.\ Paulson School of Engineering and Applied Sciences, Harvard University, 29 Oxford Street, Cambridge, MA 02138, USA}

\author{Edward Bielejec}
\affiliation{Sandia National Laboratories, Albuquerque, NM 87185, USA}

\author{Matthew Markham}
\affiliation{Element Six Global Innovation Centre, Fermi Avenue, Harwell Oxford, Didcot, Oxfordshire OX110QR, United Kingdom}

\author{Marko Lon\v{c}ar}
\affiliation{John A.\ Paulson School of Engineering and Applied Sciences, Harvard University, 29 Oxford Street, Cambridge, MA 02138, USA}


\begin{abstract}
  Solid-state single-photon emitters are attractive for realization of integrated quantum systems due to their experimental convenience and scalability. Unfortunately, however, their complex mesoscopic environments cause photons from different emitters to be spectrally distinguishable.
  Here we demonstrate spectral alignment of two solid-state single-photon emitters by utilizing the strain gradient.
  Multiple germanium vacancy (GeV) color centers in diamond are created in fabricated diamond microcantilevers using focused ion beam implantation.
  The strain response of the electronic energy levels of the GeV is measured by inducing an electrically controlled strain in the device.
  Leveraging the large strain gradient, we tune two GeVs in the spot to emit at the same optical wavelength.
  Simultaneous resonant excitation of two spectrally aligned GeVs is demonstrated using a narrow linewidth laser.
  This strain gradient tuning can be used to realize spectrally identical quantum emitters for on-chip integrated quantum systems.
\end{abstract}

\maketitle

Stable single-photon emitters~\cite{Aharonovich2016NatPhoton, Kimble1977PRL} are essential components of quantum electrodynamics experiments \cite{Bhaskar2017PRL, Pingault2017natcomm, Sipahigil2016Science, Riedrich2014NL, Scheibner2007Natphys} and optical quantum technologies \cite{Hensen2015Nature, OBrien2009NatPhoton, Kimble2008nature, OBrien2007science, Kuhn2002PRL}.
Fluorescent color centers in diamond \cite{Aharonovich2014AdvOptMat}, as well as other atom-like emitters in solids \cite{ Bracher2017PNAS, Somaschi2016natphoto, Ding2016PRL}, have emerged as a promising platform due to their outstanding optical properties and compatibility with integrated photonic approaches.
Among these, inversion-symmetric emitters such as the silicon vacancy (SiV) \cite{Pingault2017natcomm, Sipahigil2016Science, Rogers2014NatComm, Riedrich2014NL, Rose2018Science} and germanium vacancy (GeV) \cite{Iwasaki2015SciRep, Palyanov2015scirep, Siyushev2017PRB, Bhaskar2017PRL} in diamond stand out owing to their desirable optical properties such as narrow linewidth and low spectral diffusion.
However, the complex mesoscopic environments of diamond color centers lead to spectral inhomogeneities between different emitters \cite{Aharonovich2016NatPhoton}, which is a problem for many applications that rely on spectrally identical emitters.

A variety of tuning mechanisms have been used to overcome the spectral differences among solid-state emitters, including temperature \cite{Englund2007nature, Yoshie2004nature, Chen2011APL}, electromagnetic field \cite{Kim2011APL, Sipahigil2012PRL, Bernien2012PRL}, and stresses/strain \cite{Trotta2016natcomm, Chen2018arxiv, Sternschulte1994PRB, Meesala2016PRApplied}.
Local control of spectral properties of inversion-symmetric emitters \cite{Hepp2014PRL, Hepp2014Thesis} is particularly challenging since they are not sensitive to applied electrical fields, to the first order.
Mechanical strain in nanostructures provides efficient and local tuning for these emitters, and has recently been used to control the emission spectrum of SiV centers embedded in microfabricated electromechanical systems \cite{Sohn2018natcomm,Meesala2018PRB}.
In this work, we demonstrate that strong strain gradients that exist in microcantilevers can be used to spectrally align quantum emitters, with high spatial resolution.
This could be an efficient technique for overcoming inhomogeneous distribution of emitters and improving the properties of quantum networks \cite{Sipahigil2016Science, Hensen2015Nature}. 

\begin{figure}
  \includegraphics{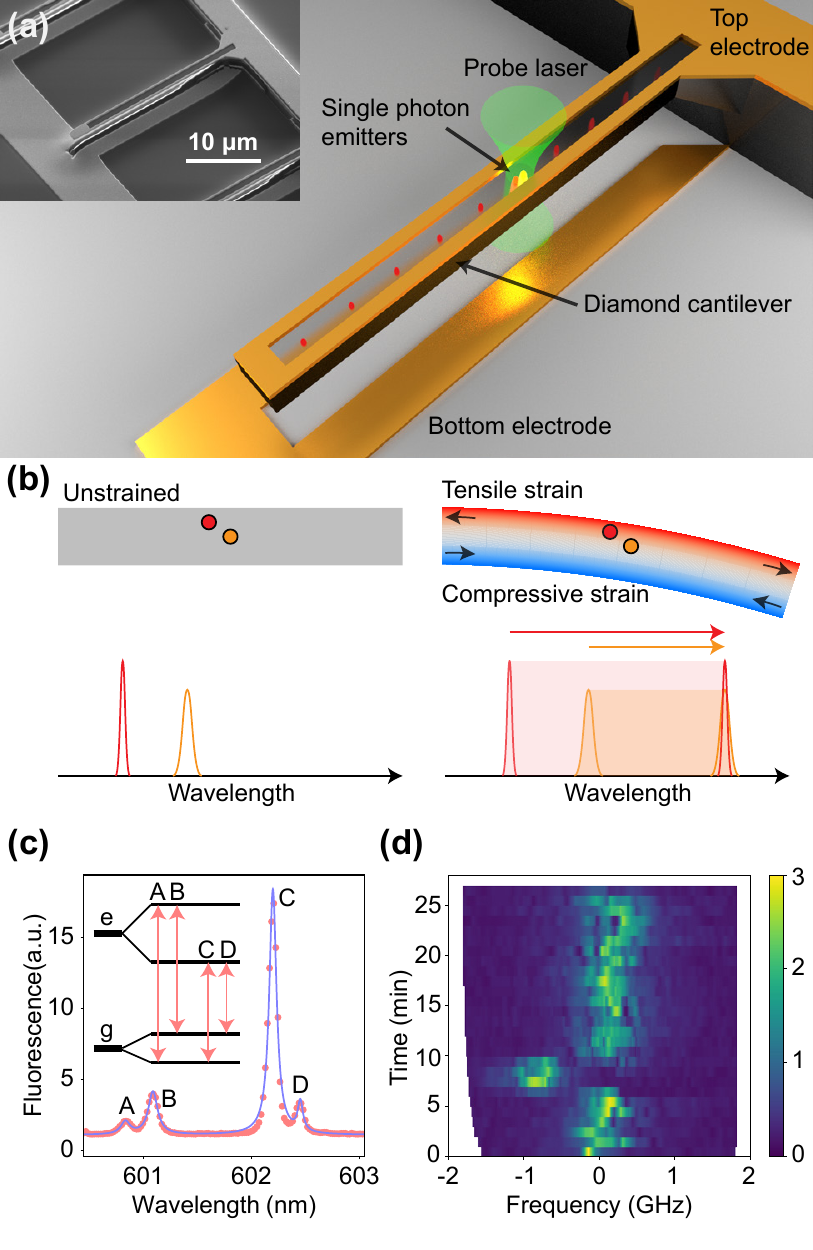}
  \caption{
    \label{fig1}
    Multiple solid-state emitters spectrally tuned by strain gradient.
    (a) Schematic of a diamond microcantilever.
    Voltage applied between the top and bottom electrodes produces an electrostatic force on the cantilever, generating strain.
    Multiple GeVs at a single site can be simultaneously accessed optically by a probe laser.
    Inset: scanning electron microscope image of a diamond microcantilever.
    (b) Illustration of spectral alignment of two emitters with the strain gradient.
    Emission spectra of the two emitters are initially different (left), and are brought in resonance by strain gradient in the deflected cantilever (right).
    (c) Photoluminescence spectrum of a GeV measured at $50\text{ K}$.
    Inset: Electronic level structure of the GeV center. g and e represent ground and excited state manifolds.
    The A-D peaks in spectrum correspond to the A-D transitions between the electronic energy levels.
    (d) Time dependent spectra of one emission line of a single GeV center in a diamond microcantilever.
    Photon counts are shown in arbitrary units.
  }
\end{figure}

A diamond microcantilever device (Fig.\ \ref{fig1}(a)) is utilized to characterize the response of the energy levels of the GeV to applied strain and achieve a tuning of over 100 GHz in the emission spectrum.
By leveraging both the large strain response of the GeV and large spatial gradient of the strain field in the device (Fig.\ \ref{fig1}(b)), we demonstrate the ability to spectrally align two GeV centers, implanted within a 50 nm spot.

The device is fabricated in a bulk diamond sample with the top surface normal to the $[001]$ crystal direction and the long axis of the cantilever along the $[110]$ direction (Fig.\ \ref{fig1}(a)).
The microcantilever pattern is realized by electron-beam lithography and then transferred into the diamond via angled ion beam etching \cite{Atikian2017APLphoton}.
The result is a suspended structure with a characteristic triangular cross sectional profile.
GeV centers are introduced deterministically into the fabricated diamond cantilevers using focused ion beam implantation ($200\text{ keV}$ $^{74}\text{Ge}^{2+}$) followed by annealing.
The lateral accuracy of the implantation is better than $50\text{ nm}$.
The resulting depth predicted by a Stopping and Range of Ions in Matter (SRIM) simulation \cite{Ziegler2010SRIM} of the implanted ions is $75\text{ nm}$ with a straggle of $12\text{ nm}$.
The electrodes are finally fabricated by electron-beam lithography and lift-off.

The GeV is a point defect in diamond consisting of one germanium atom positioned halfway between two adjacent missing carbon atoms.
It is geometrically identical to the SiV center in diamond, and has the same $D_{3d}$ symmetry. As a result, group theoretic arguments predict a qualitatively similar electronic structure for both defects \cite{Hepp2014PRL}.
The excited and ground states of the GeV are separated by about $602\text{ nm}$ in wavelength, and this transition is called the zero phonon line (ZPL). Spin-orbit coupling splits the ground (excited) state into two branches (Fig.\ \ref{fig1}(c) Inset).
To resolve individual transitions, we measure the photoluminescence spectrum of GeV centers excited by a $532\text{ nm}$ laser at temperature of $50\text{ K}$ which ensures that all energy levels are sufficiently populated.
Four emission lines in the spectrum, marked A-D in Fig.\ \ref{fig1}(c), are identified as corresponding transitions between the excited and ground states.
The C line, the transition between the lower excited state and the lower ground state, has the highest photon counts, and is the most useful for quantum optics experiments.
According to a Lorentzian fit, the center wavelengths of the emission lines A-D are at $600.837(8)\text{ nm}$, $601.091(2)\text{ nm}$, $602.2020(4)\text{ nm}$, and $602.456(3)\text{ nm}$, indicating a ground to excited state energy gap of $601.647\text{ nm}$, ground state splitting of $212\text{ GHz}$, and excited state splitting of $1138\text{ GHz}$.  
These splittings include the effect of pre-strain in the fabricated nanostructure. 
The inversion symmetry of the GeV makes it robust against external electric field fluctuations to first order, resulting in a relatively stable optical transition.
The measured spectra show very little frequency variation over tens of minutes despite the occasional spectral jumps (Fig.\ \ref{fig1}(d)), possibly due to strain or second order effects of electric field fluctuations in the local environment.

\begin{figure*}
  \includegraphics{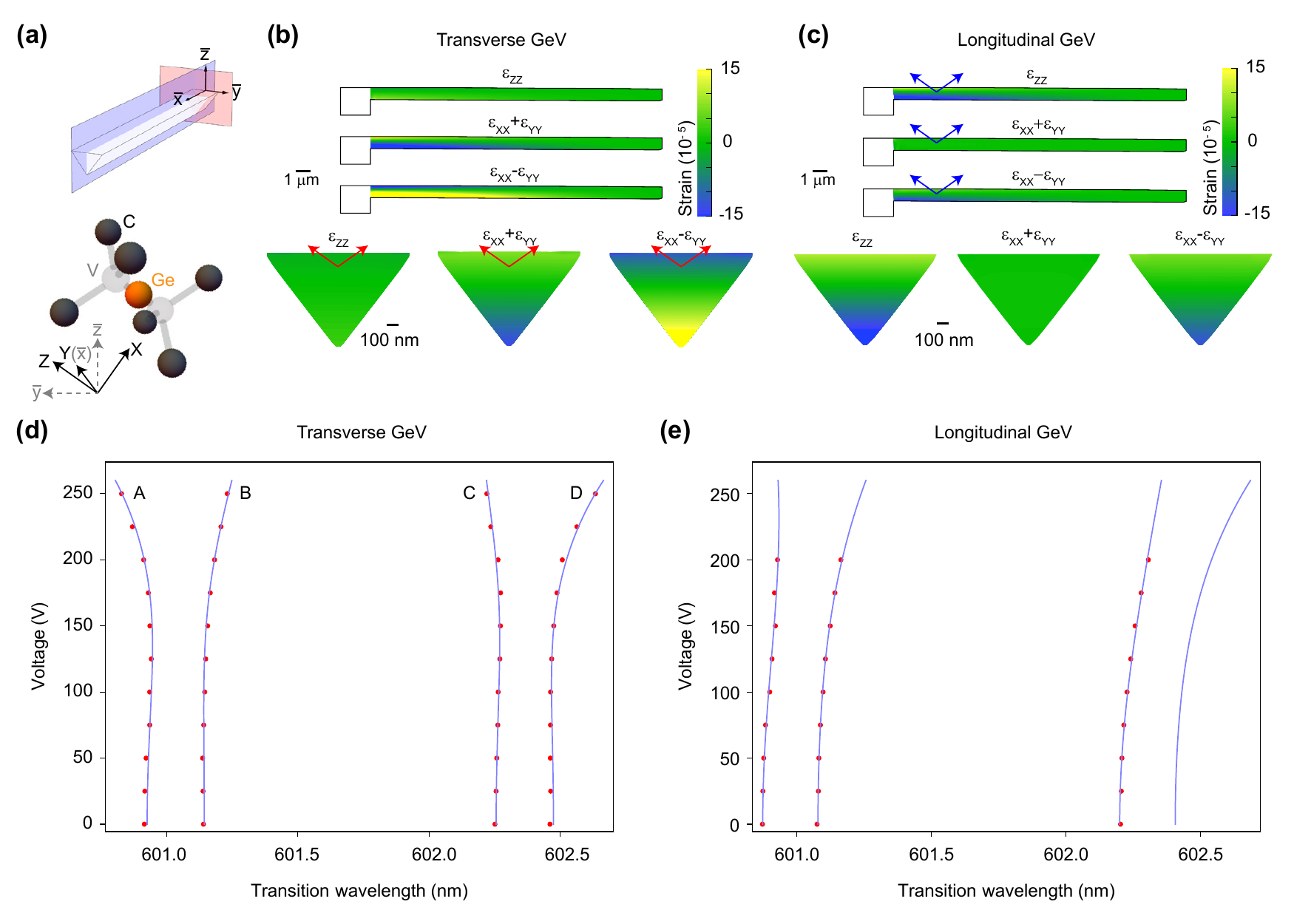}
  \caption{\label{fig2}
    Strain profiles of the diamond microcantilever and strain tuning of GeV centers.
    (a) Illustrations of the cantilever and the molecular structure of the GeV and their corresponding coordinate systems.
    $\bar{x}\bar{y}\bar{z}$ is the coordinate system of the cantilever device, and $\text{XYZ}$ is the coordinate system of the GeV.
    Ge, C, and V represent germanium, carbon, and vacancy.
    (b),(c) Simulated profiles of diagonal strain components for the transverse (b) and longitudinal (c) GeV centers with highest symmetric axis $Z$ lying in the triangular transverse cross section, and the $Z$ axis of GeV is indicated by the arrows.
    $\epsilon_\text{XX}$, $\epsilon_\text{YY}$, and $\epsilon_\text{ZZ}$ are diagonal strain components in the GeV coordinate system.
    The cantilevers are $20 \mathrm{\,\mu{m}}$, $1.28 \mathrm{\,\mu{m}}$, and $0.83 \mathrm{\,\mu{m}}$ in length, width, and height.
    The longitudinal cross sections are $\bar{x}\bar{z}$-planes at $\bar{y}=0$, and the transverse cross sections are  $\bar{y}\bar{z}$-planes at $\bar{x}=2.5 \mathrm{\,\mu{m}}$.
    (d),(e) Tuning of the spectral peaks of the transverse (d) and longitudinal (e) GeV with voltage applied to the electrodes on the microcantilever.
    The theoretical model (blue lines) is a fit to the experimental measurements (red dots).
    Only the A,B,C lines were observed experimentally for the longitudinal GeV center.
  }
\end{figure*}

Finite element simulations \cite{COMSOL} are used to evaluate the amount of strain induced at the location of the GeVs, in response to the applied voltage.
We express the strain with respect to the local coordinates of the GeV center (Fig.\ \ref{fig2}(a) Inset). 
The major axis of the GeV (the line joining the two missing carbon atoms), named the Z axis, has four possible orientations along the four equivalent $\langle 111\rangle$ diamond crystal directions.
The four orientations are grouped into two classes depending on the relative direction to the cantilever -- transverse GeVs with their major axis lying in the transverse cross section plane (Fig.\ \ref{fig2}(b)), and longitudinal GeVs with their major axis lying in the longitudinal cross section plane (Fig.\ \ref{fig2}(c)).
The two classes show qualitatively different strain responses.
Fig.\ \ref{fig2}(a) shows the relation between the cantilever and local coordinate systems for the transverse GeV.
For transverse GeVs, the diagonal elements of the strain tensor are shown in Fig.\ \ref{fig2}(b); the shear elements are at least one order of magnitude smaller and thus have negligible contribution to the overall strain response.
Since the maximum strain in the cantilever is along the $\bar{x}$ direction in the cantilever coordinate system, the strain component $\epsilon_\text{ZZ}$ for transverse GeVs is small relative to the diagonal strains $\epsilon_\text{XX}$ and $\epsilon_\text{YY}$.
Longitudinal GeVs are rotated by $90^\circ$ with respect to transverse GeVs, so that their $\text Y$ axis is aligned with the $\bar y$ axis of the cantilever instead.
Most of the strain for longitudinal GeVs is in the $\epsilon_\text{ZZ}$ component.

The strain terms $\epsilon_\text{ZZ}$ and $\epsilon_\text{XX}+\epsilon_\text{YY}$ shift the energy levels of the GeV, while the $\epsilon_\text{XX}-\epsilon_\text{YY}$ term mixes the two orbital branches within both the ground and excited manifolds \cite{Hepp2014Thesis}.
The strain tuning of the spectrum of single transverse and longitudinal GeV centers are experimentally characterized in Figs.\ \ref{fig2}(d) and (e) respectively.
The spectral lines, which have a linewidth of about 1 GHz as discussed later, are tuned by several tens of linewidths.   
For the transverse GeV, with the maximum voltage of $250\text{ V}$ applied to the cantilever electrodes, the D line shows the largest shift of $-144 \text{ GHz}$, while the bright C line shifts by $26\text{ GHz}$.
The maximum voltage that can be applied to the device is limited by the leakage current.
For the longitudinal GeV, the C line shifts by $-86\text{ GHz}$ for the maximum applied voltage of $200\text{ V}$.
The D line was not observed clearly for this GeV, due to low fluorescence intensity.

When the applied voltage is ramped up, the resulting increase of the splitting between A and B (equivalently, C and D) lines implies the corresponding increase in the energy splitting in the ground state manifold.
Similar observations for A and C (B and D) lines indicate the increasing splitting in the excited state manifold.
Fundamentally, strain affects the electronic energy levels of the GeV by deforming the orbital states.
The strain interaction modifies the electronic levels of the GeV in addition to the spin-orbit and Jahn-Teller interactions \cite{Hepp2014PRL, Hepp2014Thesis, Meesala2018PRB, Sohn2018natcomm}.
By diagonalizing the Hamiltonian of orbital states of the GeV center \cite{Supp}, we obtain the electronic energy levels given by
\begin{equation}
E_{g,e\pm} = \alpha_{g,e}\pm\sqrt{\left(\frac{\lambda_{\text{SO}\,g,e}}{2}\right)^2 + \beta_{g,e}^2 + \gamma_{g,e}^2}
\end{equation}
where $\lambda_{\text{SO}g,e}$ are the spin-orbit coupling strengths; $g$ and $e$ indicate ground and excited states, respectively.
$\alpha$, $\beta$, and $\gamma$ describe the response of the GeV electronic ground or excited states to strain, which are given by 
\begin{align}
\alpha_{g,e} &=
t_{\perp g,e} (\epsilon_\text{XX} + \epsilon_\text{YY}) + t_{\parallel g,e} \epsilon_\text{ZZ}\\
\beta_{g,e} &=
d_{g,e}(\epsilon_\text{XX} - \epsilon_\text{YY}) + f_{g,e} \epsilon_\text{ZX}\\
\gamma_{g,e} &=
-2\,d_{g,e} \epsilon_\text{XY} + f_{g,e} \epsilon_\text{YZ}
\end{align}
where $\epsilon_{ij}$ are the components of the strain tensor given in the GeV center coordinate system shown in Figs.\ \ref{fig2}(a) and S1.
$t_\perp$, $t_\parallel$, $d$ and $f$ are the four strain susceptibility parameters describing the strain response of the GeV orbital states.
The shift of the C line under strain is given by $\Delta \nu_C = E_{e,-} - E_{g,-}$.
In the limit of small strain, i.e.\ $\beta, \gamma \ll \lambda_\text{SO}$, the shift of the C line is $\Delta \nu_C \approx (t_{\parallel e}-t_{\parallel g}) \epsilon_\text{ZZ} + (t_{\perp e}-t_{\perp g}) (\epsilon_\text{XX}+\epsilon_\text{YY})$.
For large strains, the ground (excited) splitting terms $\sqrt{\lambda_\text{SO}^2/4 + d^2(\epsilon_\text{XX} - \epsilon_\text{YY})^2}$ also shift the C line significantly.
The theoretical model is consistent with the experimental measurements with fitted strain susceptibilities (Figs.\ \ref{fig2}(d) and (e)).

With the spectra of both transverse and longitudinal GeVs under various strains, a fit to the above model gives spin-orbit couplings $\lambda_{\text{SO},g} = 165\text{ GHz}$, $\lambda_{\text{SO},e} = 1098\text{ GHz}$, and an initial pre-strain of about $2\times 10^{-5}$ in the $\epsilon_\text{XX}-\epsilon_\text{YY}$ term for the transverse GeV in Fig.\ \ref{fig2}(d).
This is not surprising for color centers in fabricated nanostructures with thin gold layers deposited nearby.
The strain susceptibility coefficients obtained from the fit are $t_{\parallel e} - t_{\parallel g} = -1.7 \text{ PHz/strain}$, and $t_{\perp e} - t_{\perp g} = -0.9\text{ PHz/strain}$ which are similar to that of the SiV\cite{Sohn2018natcomm,Meesala2018PRB}.
Further by fitting the ground and excited splitting, we estimate $d_e = 2.8 \text{ PHz/strain}$ and $d_g = 2.2\text{ PHz/strain}$.
The accuracy of the estimated strain susceptibility coefficients is limited by the straggle of the implanted ion locations, as well as the large strain gradient in the device.

\begin{figure}[ht]
  \includegraphics{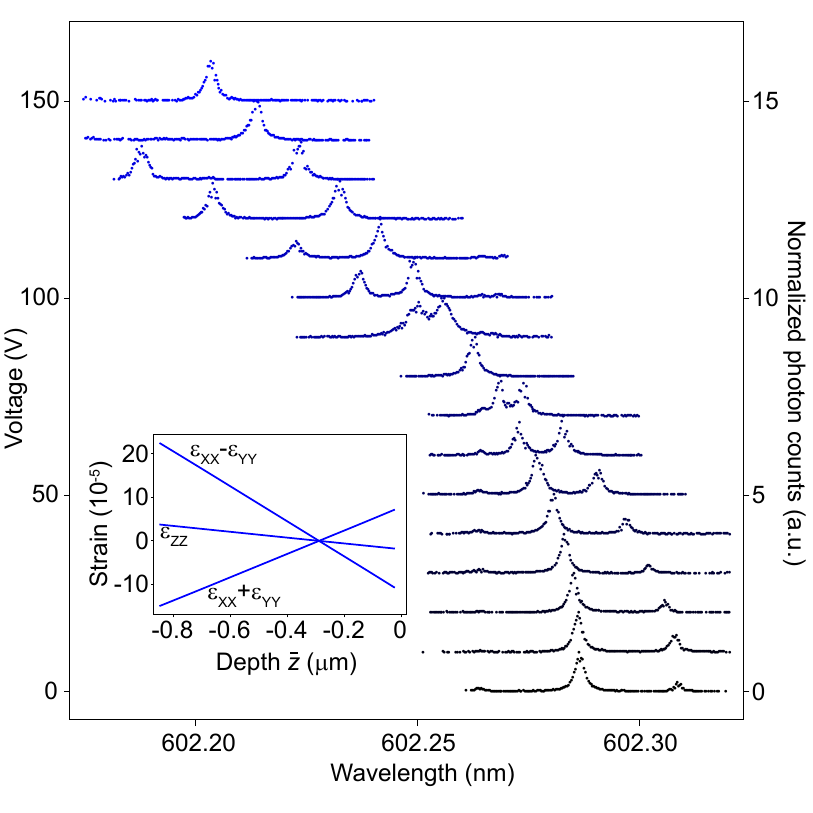}
  \caption{
    \label{fig3}
    Resonant excitation spectra showing the C transitions of two transverse GeV centers crossing as increasing voltage is applied to bend the microcantilever.
    The spectra are represented by a recorded trace of photon counts from the longer-wavelength phonon side band as a narrow linewidth laser scans its wavelength across the transitions.
    The spectra are shifted by the control voltage and normalized individually.
    Inset: $\epsilon_\text{ZZ}$, $\epsilon_\text{XX}+\epsilon_\text{YY}$, and $\epsilon_\text{XX}-\epsilon_\text{YY}$ components for transverse GeVs at various depths $\bar{z}$ at $\bar{x}=2.5 \mathrm{\,\mu{m}}$, $\bar{y}=0$.
  }
\end{figure}

When deflected, the microcantilever has a large strain gradient in the $\bar z$ direction (Fig. \ref{fig3} Inset).
As a result, two GeVs separated by tens of nanometers in depth experience a significant difference in strain, and can be tuned together spectrally with an appropriate voltage applied to the cantilever.
Interestingly, there is a neutral depth in the cantilever such that GeVs above or below this depth will experience tensile or compressive strain, respectively.
Hence, it is possible to tune the optical spectra of multiple GeVs in opposite directions.

We experimentally demonstrate spectral alignment of two GeVs using the strain gradient (Fig.\ \ref{fig3}), which is controlled by the voltage applied between the electrodes on the microcantilever and the substrate.
To enhance the resolution of the GeV spectra, we excite the GeV transitions resonantly and count the photons emitted in the longer-wavelength phonon side band (PSB).
When the probe laser is resonant with the GeV transitions, increased photon counts from the PSB are expected.
Here, we have the C lines of two GeVs initially separated by $18.5\text{ GHz}$.
With increasing voltage, the C lines of the two GeVs get closer and overlap at a control voltage of about  $80\text{ V}$, crossing at higher voltages.
The same shift direction of the two GeVs indicates that they are most likely in the same orientation class, i.e.\ transverse GeVs.
From $0\text{ V}$ to $130\text{ V}$, one GeV shifts by $52.8\text{ GHz}$, while the other one shifts by $100.6\text{ GHz}$.
In this way, the strain gradient can bring GeVs of the same orientation class into spectral overlap, even if they are initially separated by tens of GHz.

\begin{figure*}
  \includegraphics{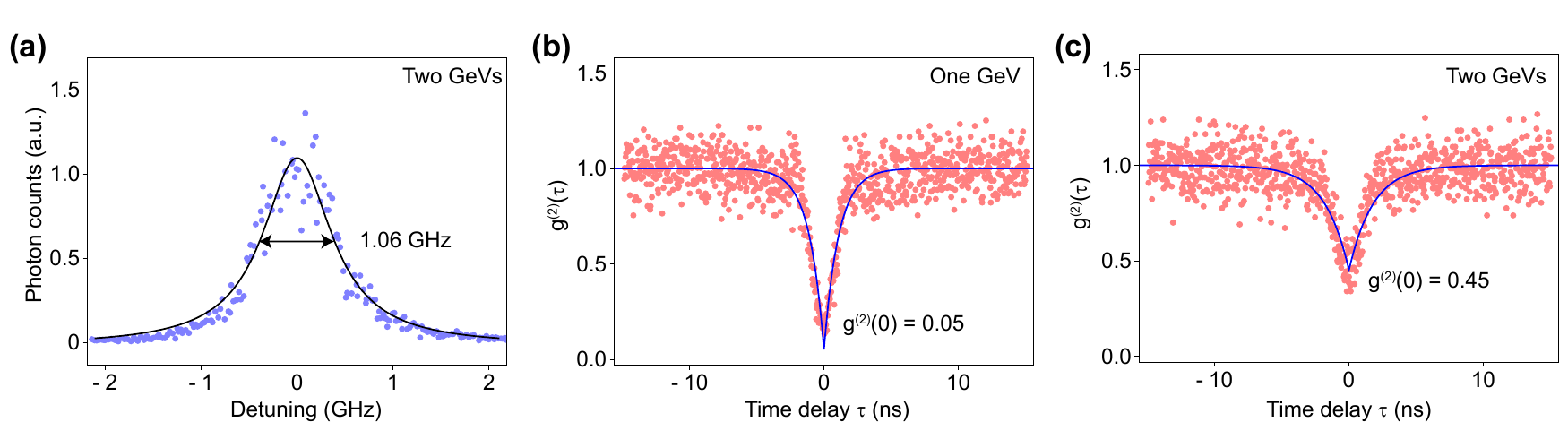}
  \caption{
    \label{fig4}
    (a) Resonant excitation spectrum of the two overlapping C transitions of the two GeV centers.
    (b) Second-order autocorrelation $g^{(2)}(\tau)$ of the PSB photons emitted from one GeV on resonant excitation when the two GeVs are tuned apart.
    (c) Second-order autocorrelation $g^{(2)}(\tau)$ of the PSB photons on resonant excitation when the two GeVs are tuned together.
    The red dots in (b) and (c) are the raw data of the event counts in the second-order autocorrelation measurements.
    The blue curves show the theoretical fit.
  }
\end{figure*}

We further investigate the extent of spectral overlap of the two GeVs by measuring correlations between photons emitted into the PSB on resonant excitation of the C transitions, as shown in Fig.\ \ref{fig4}.
We split the collected photons into two paths; the arrival of a photon in one path triggers a timer that stops upon detection of a photon in the other path.
This measurement corresponds to the second-order autocorrelation function $g^{(2)}(\tau) = \langle I(t+\tau) I(t)\rangle / \langle I(t+\tau)\rangle \langle I(t)\rangle$, where $I(t)$ and $I(t+\tau)$ are the intensities of PSB photon streams at time $t$ and a delayed time $t+\tau$, respectively.
When the two GeVs are spectrally distinguishable, we can only resonantly excite one GeV at a time.
The fitted value of $g^{(2)}(0) = 0.05\pm 0.03$ indicates a single photon source (Fig.\ \ref{fig4}(b)).
Starting from the previous voltage of $80\text{ V}$ for spectral overlap, we finely tune the voltage between the cantilever electrodes to minimize the inhomogeneous distribution of the resonant excitation spectrum of the two GeVs (Fig.\ \ref{fig4}(a)).
When the two GeVs are resonantly excited simultaneously, the fitted $g^{(2)}(0) = 0.45\pm 0.03$ (Fig.\ \ref{fig4}(c)).
The $g^{(2)}(0)$ differs from the theoretically expected value of $0.5$ for two single-photon emitters, which could be a result of different excitation rates, different collection efficiencies, or different decay rates for the two GeVs.

In conclusion, the diamond nanoelectromechanical system technique is utilized to optically characterize the strain response of GeV color centers --- an emerging single photon source in diamond.
It provides a way to optically image the strain in diamond nanomechanical systems.  
Furthermore, strain gradient in an electrically deflected microcantilever is used to overcome the inhomogeneous distribution of different GeVs which renders them spectrally distinguishable.
Importantly, our technique has spatial selectivity comparable to that of B-field gradient method developed for diamond NV centers \cite{Chen2013NL}.
As the strain gradient tensor provides sufficient degrees of freedom, advanced nanoelectromechanical devices with multiple electrodes could be utilized to simultaneously spectrally align multiple emitters.
Furthermore, our platform may enable in situ manipulation and entanglement of multiple solid-state qubits.
The ability to switch the spectral indistinguishability of quantum emitters in the same excitation volume can enable controllable local interactions and entanglement between spin qubits, and increase quantum memory density by having multiple qubits at a single spot that can be individually addressed through a single optical path.
This approach can be scaled to spectrally align emitters in multiple cantilevers simultaneously, by adjusting the voltage on each cantilever appropriately.

\begin{acknowledgments}
  
  The authors thank Dr.~Mian Zhang, Dr.~Cheng Wang, Dr.~Stefan Bogdanovic, Dr.~Michael Burek, Dr.~Vivek Venkataraman, Dr.~Alp Sipahigil, Cleaven Chia, and Mihir Bhaskar for fruitful discussions, and Daniel L.\ Perry for the focused ion implantation.
  This work is supported by the Center for Integrated Quantum Materials (NSF grant No.\ DMR-1231319), ONR MURI on Quantum Optomechanics (Grant No.\ N00014-15-1-2761), NSF EFRI ACQUIRE (Grant No.\ 5710004174), NSF GOALI (Grant No.\ 1507508), and ARL CDQI (Grant No.\ W911NF1520067).
  Focused ion beam implantation was performed under the Laboratory Directed Research and Development Program at the Center for Integrated Nanotechnologies, an Office of Science User Facility operated for the U.S.\ Department of Energy (DOE) Office of Science.
  Sandia National Laboratories is a multi-mission laboratory managed and operated by National Technology and Engineering Solutions of Sandia, LLC., a wholly owned subsidiary of Honeywell International, Inc., for the U.S.\ Department of Energy's National Nuclear Security Administration under contract DE-NA-0003525.

  S.Maity\ and L.S.\ performed experimental measurements.
  L.S.\ and Y.-I.S.\ fabricated the devices.
  S.Maity\ and L.S.\ built the experimental setups with discussions from S.Meesala.
  E.B.\ performed the focused ion beam implantation.
  S.Maity\ and L.S.\ prepared the manuscript with discussions from all authors.
  M.L.\ supervised this project.
\end{acknowledgments}


%

\end{document}